\documentclass[aps,twocolumn,raggedbottom,prb,showpacs,nobalancelastpage,amssymb,superscriptaddress]{revtex4}
\usepackage[latin1]{inputenc}
\usepackage[english]{babel}
\usepackage{graphicx}
\usepackage{psfrag}
\usepackage{amssymb}
\usepackage{amsmath}
\usepackage{amscd}
\usepackage{eucal}
\usepackage{color}
\usepackage{bm}
\usepackage[normalem]{ulem}

\newcommand{\bi}{\bibitem}

\newcommand{\rmi}{{\rm i}}

\newcommand{\half}{{\textstyle{\frac{1}{2}}}}

\newcommand{\up}{\uparrow}
\newcommand{\dn}{\downarrow}

\newcommand{\al}{\alpha}
\newcommand{\be}{\beta}

\newcommand{\eps}{\epsilon}
\newcommand{\veps}{\varepsilon}

\begin{document}

\title{
Onsager Relations in Coupled Electric, Thermoelectric and Spin Transport: \\
The Ten-Fold Way}
\author{Philippe Jacquod}
\affiliation{Physics Department, University of Arizona,
Tucson, AZ 85721, USA}
\affiliation{Theoretical Physics Department, University of Geneva, 1211 Geneva, Switzerland}
\author{Robert S.~Whitney}
\affiliation{
Laboratoire de Physique et Mod\'elisation des Milieux Condens\'es (UMR 5493), 
Universit\'e Grenoble 1, Maison des Magist\`eres, B.P.~166, 38042 Grenoble, France.}
\author{Jonathan Meair}
\affiliation{Physics Department, University of Arizona,
Tucson, AZ 85721, USA}
\author{Markus B\"uttiker}
\affiliation{Theoretical Physics Department, University of Geneva, 1211 Geneva, Switzerland}
\date{September 19, 2012}
\begin{abstract}
Hamiltonian systems can be classified into ten classes, in terms of the presence or absence of 
time-reversal symmetry, particle-hole symmetry and sublattice/chiral symmetry. 
We construct a quantum coherent scattering theory of linear transport for coupled electric, 
heat and spin transport; including the effect of Andreev reflection from superconductors.
We derive a complete list of the Onsager reciprocity relations between
transport coefficients for coupled electric, spin, thermoelectric and spin caloritronic effects.
We apply these to all ten symmetry classes, paying special attention to specific additional
relations that follow from the combination of symmetries, beyond microreversibility.
We discuss these relations in several illustrative situations. 
We show the reciprocity between spin-Hall and inverse spin-Hall effects,
and the reciprocity between spin-injection and magnetoelectric spin currents.
We discuss the symmetry and reciprocity relations of Seebeck, Peltier, spin-Seebeck and spin-Peltier
effects in 
systems with and without coupling to superconductors.
\end{abstract}
\pacs{73.23.-b,85.75.-d,72.15.Jf,74.25.fg}
\maketitle

\section{Introduction}

Onsager's reciprocity relations are cornerstones of nonequilibrium statistical 
mechanics~\cite{Onsager1,Onsager2}. They relate
linear response coefficients between flux densities and thermodynamic forces
to one another. They are based on the fundamental principle of
microreversibility which, for systems with time-reversal symmetry (TRS), says that
``if the velocities of all the particles present
are reversed simultaneously the particles will retrace their former paths,
reversing the entire succession of configurations''~\cite{Onsager1}.
When TRS is broken, microreversibility further
requires to invert all TRS breaking fields, which, to fix ideas, one may take
as magnetic fields, fluxes or exchange fields. Combining all of them into a single multi-component 
field ${\cal H}$, the Onsager reciprocity relations
read~\cite{Onsager1,Onsager2}
\begin{equation}\label{eq:Onsager_original}
{\cal L}_{ij}({\cal H})={\cal L}_{ji}(-{\cal H}) \, ,
\end{equation}
where the 
linear coefficient ${\cal L}_{ij}$ determines the response of the 
flux density ${\cal J}_{i}$ -- for instance the electric or heat current -- 
to a weak thermodynamic force ${\cal X}_j$ -- for instance an electric field or a
temperature gradient. 
Thus the precise form of the Onsager reciprocity relations depends on the symmetries
of the system. 
Seminal works have classified noninteracting quantum mechanical systems into 
ten general symmetry classes~\cite{dys62,ver93,alt97}, and it is the purpose
of the present manuscript  to derive Onsager's relations for all these
symmetry classes. 
Four of them, in particular, combine two 
different types of quasiparticles~\cite{alt97}, with microscopic representations
including, e.g. hybrid systems where
quantum coherent normal metallic conductors are connected to superconductors.
Since at sub-gap energies an interface between a normal metal and a superconductor blocks 
heat currents but not electric currents ~\cite{And64}, it is
natural to ask whether the Onsager
reciprocity relation between, say, the Seebeck and Peltier thermoelectric coefficients
survives in such systems. 
Onsager relations in the presence of superconductivity
have been discussed rather incompletely until now~\cite{Cla96,Vir04,Tit08,Jac10,Eng11},
despite much experimental~\cite{Eom98,Cad09,Par03} and 
theoretical~\cite{Cla96,Vir04,Tit08,Jac10,Eng11,Sev00,Bez03} interest in thermoelectric transport
properties of hybrid normal-metallic/superconducting
systems. 

\begin{table*}[t]
\begin{tabular}{| l  |  l || c | c | c || c | } 
\hline
\multicolumn{2}{|c||}{{\bf Symmetry class}}   & \ {\bf TRS} \ &  \ {\bf PHS} \ &\  {\bf SLS}\  & {\bf Physical example} \\ 
\hline \hline
Wigner-Dyson & \ A (unitary) & 0 & 0 & 0 & mag.~flux $\Big.$ (no SC) \\
\  & \ AI (orthog.) & $+1$ & 0& 0 & no mag.~flux \& no spin-orbit $\Big.$ (no SC)\\
& \ AII (sympl.) & $-1$ &  0& 0 &   spin-orbit \& no mag.~flux$\Big.$ (no SC)\\
\hline \hline
Chiral & \ AIII (unitary) & 0 & 0 & 1 & mag.~flux \& bipartite lattice $\Big.$ (no SC) \\
& \ BDI (orthog.) & $+1$ & +1 & 1 & no mag.~flux \& no spin-orbit \& bipartite lattice $\Big.$ (no SC)\\
 & \ CII (sympl.) & $-1$ &  $-1$ & 1 &   spin-orbit  \& no mag.~flux \& bipartite lattice $\Big.$ (no SC)\\
\hline \hline
Altland-Zirnbauer\  
&\ D& 0 & $+1$ & 0 &  SC, mag.~flux, \& spin-orbit \\
\  
&\ C    &   0     & $-1$  & 0 &  SC, mag.~flux, \& no spin-orbit \\
&\ DIII & $-1$  & $+1$ & 1 &  SC, no mag.~flux, \& spin-orbit\\
&\ CI   &  $+1$ & $-1$ & 1  & SC, no mag.~flux,  \& no spin-orbit\\
\hline 
\end{tabular}
\caption{\label{table1}
The ten-fold symmetry classification of Hamiltonians. 
The second column from the left refers to Cartan's nomenclature
for symmetric spaces~\cite{alt97}, the three middle columns indicate whether
the classes have broken (0) or unbroken ($\pm 1$)
time-reversal symmetry (TRS), 
particle-hole symmetry (PHS) and sublattice symmetry (SLS), 
while the rightmost column mentions microscopic realizations in each class,
with SC indicating the presence of superconductivity.
Aside from the indicated bipartite lattice models with preserved sublattice symmetry, 
the chiral classes are also realized in low-energy models for quantum chromodynamics~\cite{ver93}.
Spin rotational symmetry (SRS) is present in classes AI, BDI, C and CI,
absent in classes AII, CII, D and DIII, and irrelevant in classes A and AIII.
}
\end{table*}

Further motivation is provided by fundamental aspects of spintronics~\cite{Fab07} 
and spin caloritronics (spin-Seebeck and spin-Peltier effects)~\cite{bau11}, where Onsager relations 
are of significant interest~\cite{han05,sas07,tse08,bau10,Brataas1,Brataas2}.
As a matter of fact, reciprocity relations
decisively helped 
in experimentally uncovering elusive spin effects, by suggesting 
to measure electric effects that are reciprocal to them.  As but one example, we mention the inverse spin Hall 
effect~\cite{val06,Sai06,Kim07,Sek08,Liu11}, where a transverse electric current or voltage is generated by an injected
spin current~\cite{Gorini}. 
Onsager relations also put constraints on the measurement of spin currents~\cite{ada06,stano1}
that can be circumvented in the nonlinear regime only~\cite{vanwees,stano2}.
Accordingly, we will incorporate spin currents and accumulations into our 
formalism. Several of the Onsager relations for spin transport we derive below
appeared in one way or another in earlier publications, see in particular 
Refs.~[\onlinecite{sas07,tse08,bau10,Brataas1,Brataas2,val06,ada06}]. Here, we summarize them in a 
unified way and extend them to all ten symmetry
classes. We are unaware of earlier discussions of Onsager relations for
spin transport in the presence of superconductivity.

The classification into ten different symmetry classes has recently received renewed
attention, because the existence of topologically nontrivial phases~\cite{Qi11} depends on 
the system's symmetries and its dimensionality~\cite{sch08}. 
The Onsager relations we derive below depend only on fundamental symmetries
and are equally valid in topologically trivial and nontrivial states~\cite{caveat1}. 
In several instances, however,
specific additional relations exist, that arise because of 
the conservation of each quasiparticle species 
(relevant to systems without superconductor so that there are no Andreev processes converting electrons into holes and vice-versa),
the presence of particle-hole symmetry or sublattice/chiral symmetry. 
This is the case, for instance, for two-terminal 
thermoelectric transport in the Wigner-Dyson symmetry classes ~\cite{dys62}, where the relation
between Seebeck, $B$ and Peltier, $\Gamma$ coefficients reads equivalently
$B({\cal H}) T_0=\Gamma({\cal H})$ or $B({\cal H}) T_0=\Gamma(-{\cal H})$, with the base temperature
$T_0$, because of the 
additional reciprocity relation $\Gamma({\cal H})=\Gamma(-{\cal H})$~\cite{But90}. Below, 
we pay special attention to these  nongeneric relations.

The manuscript is organized as follows. In Section~\ref{sec:10fold}, we discuss the
ten symmetry classes, and the crossover between 
Altland-Zirnbauer and the Wigner-Dyson classes as the temperature is raised in hybrid
systems. In Section~\ref{sec:scat_sym} we derive and list the symmetries that the ${\cal S}$-matrix 
satisfies in all classes. Onsager relations will follow from these symmetries, once they are
inserted into scattering theory expressions for the linear transport coefficients.
In Section~\ref{sec:scat}, we formulate the problem in
terms of the scattering matrix of the system and connect the Onsager coefficients to the 
system's scattering matrix. 
In Section~\ref{sec:onsager} we list the general reciprocity relations and mention additional
ones occurring in special circumstances. Finally, in Sections~\ref{sec:examples-spin}
and \ref{sec:examples-sc}, we 
discuss some cases of importance and the associated Onsager relations in systems with 
coupled electric and spin transport, superconductors or chiral symmetries.
Conclusions are given in Section~\ref{sec:conclusion}

\begin{table*}
\begin{tabular}{| l  |  l || c | c | c || c | } 
\hline
\multicolumn{2}{|c||}{{\bf Symmetry class}}   & \ {\bf TRS} \ &  \ {\bf PHS} \ &\  {\bf SLS}\  & {\bf Physical example} \\ 
\hline \hline
{\bf Crossovers}   &  \ D $\to$ A      &  0 &  \ $+1\to 0$ \  & 0&     as D but $(\tau_{\rm Andr}k_{\rm B}T_0)$ is not small \\
\ for  Andreev    \ \    & \ C $\to$ A  &  $0$ &   \ $-1\to 0$ \  & 0 &  as C but$(\tau_{\rm Andr}k_{\rm B}T_0)$ is  not small\ \\
\  interfero.        & \ DIII $\to$ AII  &  $-1$ &   \ $+1\to 0$ \  &  1 $\to$ 0 &  as DIII but $(\tau_{\rm Andr}k_{\rm B}T_0)$ is  not small\ \\
         & \ CI $\to$ AI         & $+1$ &     \ $-1\to0$ \ & 1$\to$ 0  & as CI but $(\tau_{\rm Andr}k_{\rm B}T_0)$ is  not small \ \\
\hline
\end{tabular}
\caption{\label{table2}
Crossovers from the Altland-Zirnbauer to the Wigner-Dyson classes. Increasing the temperature
breaks PHS so that the quasiparticle excitation energy $\eps$ cannot be treated as small.
One way to break PHS is to make $(k_{\rm B}T_0)\tau_{\rm Andr}$ not negligible,
where $\tau_{\rm Andr}$ is a timescale associated with impinging on or returning to the superconducting
contacts.
}
\end{table*}

\section{The ten-fold way}\label{sec:10fold}

Hamiltonian systems are classified according to the presence or absence of
fundamental symmetries.
The historical classification scheme~\cite{dys62,alt97} 
is based on TRS and  spin-rotational symmetry (SRS). Three {\it Wigner-Dyson} 
classes are defined in 
this way. Using the Cartan nomenclature for symmetric spaces, the class A has both
symmetries broken, the class AI has both symmetries present, and the class AII
has broken SRS but unbroken TRS. When
TRS is broken, the presence or
absence of SRS only affects the size of the Hamiltonian matrix --- and not its symmetry --- 
and there is thus no fourth class.

Chiral classes were next introduced~\cite{ver93}, which
capture the structure of the QCD Dirac operators. Beside relativistic fermions, they are
also appropriate to describe bipartite lattice Hamiltonians with unbroken sublattice
symmetry (SLS). Examples include two-dimensional 
square and hexagonal lattices, as well as three-dimensional cubic lattices  
without mass/on-site term, the latter generically breaking SLS. 
Here also, there
are three classes, with (apart from their chiral symmetry) the same
symmetries as the Wigner-Dyson classes. 

Finally, 
four more classes of Bogoliubov-de Gennes (BdG) Hamiltonians
appear - the Altland-Zirnbauer classes~\cite{alt97} - when normal metals are
brought into contact with superconductors: with
SRS (C and CI) and without SRS (D and DIII), 
with TRS (CI and DIII) and without TRS (C and D). When dealing with such systems,
we use a convention where the 
BdG Hamiltonian reads
\begin{eqnarray}\label{eq:bdg}
H = \left(
\begin{array}{cc}
h - \mu_{\rm sc} & \Delta \\
\Delta^* & \mu_{\rm sc} - \sigma^{(y)} h^* \sigma^{(y)} 
\end{array}
\right) \, ,
\label{Eq:H-BdG}
\end{eqnarray}
with the Pauli matrix $\sigma^{(y)}$ acting on the spin degree of freedom and 
the chemical potential $\mu_{\rm sc}$ on the superconductor. 
With this convention, used for example in Refs.~[\onlinecite{Slevin1996,Oreg-Majorana-2010}], 
the second-quantized Bogoliubov-de Gennes Hamiltonian is 
$\half \boldsymbol{c}^\dagger  H \boldsymbol{c}$, where 
\begin{eqnarray}
\boldsymbol{c}^\dagger = (c_{e\uparrow}^\dagger ,c_{e\downarrow}^\dagger ,c_{h\downarrow}^\dagger ,
-c_{h\uparrow}^\dagger ).
\label{Eq:Psi-for-e-and-h}
\end{eqnarray}
with $c_{e\uparrow}^\dagger$ being the vector of creation operator for all $k$-states of spin-$\uparrow$ electrons, etc.
This has the hole sector rotated by $i \sigma^{(y)}$ with respect to the Hamiltonian in 
Refs.~[\onlinecite{alt97,sch08}]. 
The form of Eq.~(\ref{Eq:H-BdG}) has the advantage that upon assuming SRS (so $h^*$ commutes with $\sigma^{(y)}$),
it immediately reduces to that used in 
Refs.~[\onlinecite{Cla96,deGennes-book,Tinkham-book,Beenakker-review}].

Ref.~[\onlinecite{sch08}] introduced a unifying {\it ten-fold} 
classification scheme for all the above Hamiltonians.  They considered TRS and particle-hole symmetry (PHS), which 
can both be represented by antiunitary operators, and
accordingly, these two symmetries can be either broken, or unbroken. In the former case, we represent
this by a $0$, while in the latter case,  the antiunitary operator squares to either $+1$ 
or $-1$. 
A squared TRS of $+1$ corresponds to spinless or integer-spin particles, while 
a squared TRS of $-1$ corresponds to half-integer-spin particles. 
A squared PHS of $+1$ corresponds to triplet pairing, while a squared PHS
of $-1$ corresponds to singlet pairing in a Bogoliubov-de Gennes 
Hamiltonian~\cite{footnote:TRS-plus-minus-one}.
Naively one would think
that this leads to $3 \times 3 = 9$ classes, however there are two distinct possibilities 
when both TRS and PHS are 
broken. In this case, the symmetry represented by the product of the two antiunitary
operators gives either $0$ (when the corresponding symmetry is broken) or $+1$. 
This finally gives $3 \times 3 - 1 + 2 = 10$ symmetry
classes again. Using the just defined three indices, we summarize the 
ten symmetry classes in Table~\ref{table1}, where we additionally mention relevant physical
realizations for each of them.

In this classification, the possible symmetries that the Hamiltonian $H$ satisfies are
(i) TRS : $H = T H T^{-1}$, with $T=-i K$, with the complex conjugation operator $K$ in the spinless case,
and $T=-i \sigma^{(y)} K$ for spin-$1/2$ fermions, with the Pauli matrix $\sigma^{(y)}$ acting in 
spin space; (ii) PHS : $H = -P H P^{-1}$, with $P=-i \sigma^{(y)}  \tau^{(y)} K$, with the 
Pauli matrix $\tau^{(y)}$ acting in Nambu space~\cite{caveat2};
(iii) SLS : $H = -\eta^{(z)} H \eta^{(z)}$, with the Pauli matrix $\eta^{(z)}$ acting on sublattice space (bipartite
lattices are assumed here).

TRS, SRS and SLS can be broken by an orbital magnetic field, spin-orbit
interaction and mass/on-site terms respectively. 
The Altland-Zirnbauer classes assume PHS, which strictly speaking forces thermoelectric
effects to vanish identically. PHS
can be broken, for instance, by moving away in energy from the special $\epsilon=0$ symmetry point -- the
superconductor's chemical potential. This occurs upon increasing the temperature, when the latter
exceeds a Thouless energy scale, $E_{\rm T} \simeq \tau^{-1}_{\rm Andr}$ where the time scale
$\tau_{\rm Andr}$ is related to the time it takes to impinge on or return to the normal-metal/superconductor
interface. This energy is implicitly assumed to be 
much smaller than the superconductor's critical temperature.
Table~\ref{table2} summarizes the crossovers from the Altland-Zirnbauer to the Wigner-Dyson
classes as the temperature is raised such that the coherence between electron and Andreev-reflected
hole quasiparticle gets lost. We will get back to this point in Section~\ref{sec:breakingphs} below.
The existence of SLS also requires that the spectrum is symmetric about zero energy, thus, at
half-filling, SLS also leads to the vanishing of thermoelectric effects, which one recovers as the 
electrochemical potential is tuned away from half-filling.

\section{Symmetries and Reciprocities of the ${\cal S}$-matrix}\label{sec:scat_sym}

Our investigations are based on the scattering theory of quantum transport~\cite{But86,Imr86},
which, for noninteracting systems, 
allows to straightforwardly derive Onsager reciprocity relations solely from the 
symmetries of the system's scattering matrix ${\cal S}$.

\begin{figure*}
\includegraphics[width=10cm]{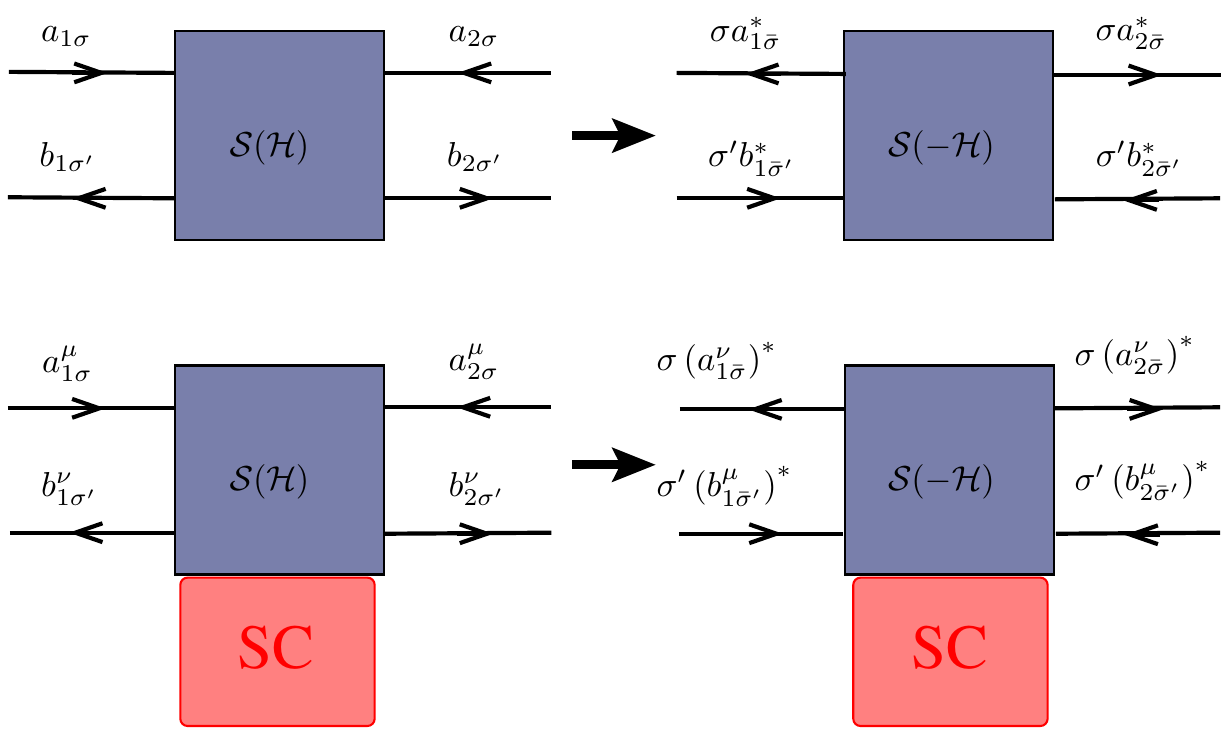}
\caption{\label{fig:microrev} Microreversibility operating on a two-terminal 
${\cal S}$-matrix in the 
absence (top) and in the presence of 
superconductivity (bottom). The scattering
amplitudes $a$ and $b$ depend on spin indices,
$\sigma=\uparrow (+1)$, 
$\downarrow (-1)$, quasiparticle indices, $\mu=e (+1)$, $h (-1)$ and terminal
indices 1 and 2.
Time-reversal implies
inverting the particle's momentum, spin and quasiparticle isospin, as well as magnetic
fields and fluxes.}
\end{figure*}

Reciprocity relations for ${\cal S}$ 
follow directly from microreversibility~\cite{But86}.
They read~\cite{zha05},
\begin{equation}\label{eq:smatr}
{\cal S} ({\cal H}) = \sigma^{(y)} \, 
{\cal S}^{\rm T}(-{\cal H}) \, \sigma^{(y)} \, , 
\end{equation}
where $\sigma^{(y)}$ is a Pauli matrix acting in spin space,
and ``$^{\rm T}$'' indicates the matrix transpose of spin, transport channel and  (with superconductivity) quasiparticle indices.
Included in Eq.~(\ref{eq:smatr}) is the relation ${\cal S}(\cal H) = {\cal S}^{\rm T}(-{\cal H})$ valid when
the antiunitary TRS operator squares to 1 and SRS is not broken.
Eq.~(\ref{eq:smatr}) can be derived
by constructing ${\cal S}$ first with scattering states $\phi_{n \sigma}({\cal H})$, then
with their
time-reversed $-i \sigma^{(y)} K \phi_{n \sigma}(-{\cal H})$, with the complex conjugation operator $K$,
and equating the two results~\cite{zha05}. Eq.~(\ref{eq:smatr}) is intimately related to Kramers degeneracy,
which in the presence of TRS (${\cal H}=0$)  follows from the symmetry property 
$H = \sigma^{(y)} H^* \sigma^{(y)}$ of the Hamiltonian $H$. Specifying to half-integer-spin
particles, 
it can  equivalently be rewritten in a form that renders its connection to
microreversibility more evident
\begin{equation}\label{eq:smatr_lms}
{\cal S}^{\mu \nu}_{i\sigma,j\sigma'}({\cal H}) = \sigma \sigma' 
{\cal S}^{\nu \mu}_{j\bar{\sigma}',i\bar{\sigma}} (-{\cal H}) \, ,
\end{equation}
where $i,j$ are transport channel indices, $\mu,\nu=e,h$ are quasiparticle
indices and $\bar{\sigma}=-\sigma$ are spin indices.

Further relations can be constructed by combining Eqs.~(\ref{eq:smatr}) and (\ref{eq:smatr_lms}) 
with additional symmetries of the ${\cal S}$-matrix.
The latter are obtained by translating PHS and SLS
of the Hamiltonian into symmetries of the ${\cal S}$-matrix. For this purpose, we use the relation~\cite{Mahaux}
\begin{equation}
{\cal S}(\eps) = 1 + 2 \pi i W^\dagger (H-\eps - i \pi W W^\dagger)^{-1} W \, , 
\end{equation}
between ${\cal S}$ and $H$,
with a rectangular matrix $W$ that couples the scatterer  to external leads.
We consider first PHS.
The presence of superconductivity requires to introduce electron and hole 
quasiparticles, and when PHS is present, the energy spectrum is
symmetric about zero energy (taken as the chemical potential of the 
superconductor). 
With the convention of Eq.~(\ref{eq:bdg}),
PHS reads
$H = -\sigma^{(y)}\tau^{(y)} H^* \sigma^{(y)} \tau^{(y)}$~\cite{caveat2b},  
with the minus-sign indicating how the symmetry of the energy spectrum differs from Kramers degeneracy. From this
we obtain~\cite{caveat3}
\begin{equation}\label{eq:smatr_sc}
{\cal S}({\cal H})  = \sigma^{(y)}\tau^{(y)} \, 
{\cal S}^*({\cal H})  \,\sigma^{(y)} \tau^{(y)} \, .
\end{equation}
Combining Eqs.~(\ref{eq:smatr}) and (\ref{eq:smatr_sc}) and 
paying attention to the ordering of spin-indices given in Eq.~(\ref{Eq:Psi-for-e-and-h}), 
one obtains, 
\begin{eqnarray}
\hskip -3mm {\cal S}^{\mu \nu}_{i \sigma, j \sigma'} ({\cal H}) \, 
&=& \,  \left({\cal S}^{\bar{\mu} \bar{\nu}}_{i \sigma, j \sigma'} ({\cal H}) \right)^* 
\nonumber \\
&=& \, 
\sigma \sigma' {\cal S}^{\nu \mu}_{j \bar{\sigma}', i \bar{\sigma}} (-{\cal H}) \, \nonumber \\
&=& \, 
\sigma \sigma' \left({\cal S}^{\bar{\nu} \bar{\mu}}_{j \bar{\sigma}', i \bar{\sigma}} (-{\cal H}) \right)^* \, . \qquad
\end{eqnarray}
where quasiparticle indices $\mu, \nu= +1(e),  -1(h)$ when they appear as prefactors.

The reciprocity relations (\ref{eq:smatr}) and (\ref{eq:smatr_sc}) for the ${\cal S}$-matrix
are illustrated in Fig.~\ref{fig:microrev}. 
We defined the ${\cal S}$-matrix via  the relation
$ {\cal S} \,
 {\bf a} = {\bf b} $
 between vectors ${\bf a}$ and ${\bf b}$  of components for incoming 
 and outgoing quasiparticle flux amplitudes, respectively. Each component of these
 vectors corresponds to 
 a given terminal, a transverse transport channel in that terminal, a spin
 orientation and, in the presence of superconductivity, a quasiparticle index.
 Complex conjugation in Fig.~\ref{fig:microrev} and Eq.(\ref{eq:smatr})
 occurs because TRS and PHS are
represented by antiunitary operators, i.e. products of 
a unitary operator with complex conjugation.

We finally comment on SLS. The chiral 
Hamiltonian symmetry reads $H = - \eta^{(z)} \, H \, \eta^{(z)}$, with the Pauli matrix $\eta^{(z)}$ acting in 
sublattice space. For the scattering matrix, this translates into 
\begin{eqnarray}\label{eq:smatr_chiral}
{\cal S}({\cal H}, \eps)   = \eta^{(z)} \, 
{\cal S}^\dagger ({\cal H},-\eps) \, \eta^{(z)} \, ,
\end{eqnarray}
where in contrast to earlier symmetry relations, 
we explicitly had to write the energy-dependence of the ${\cal S}$-matrix.
Combining Eqs.~(\ref{eq:smatr}) and (\ref{eq:smatr_chiral}), one obtains, 
\begin{eqnarray}\label{eq:smatr_chiral_rec}
 \hskip -3mm {\cal S}^{m n}_{i \sigma, j \sigma'} ({\cal H},\eps) \, 
&=& \, m n  \left({\cal S}^{n m}_{ j \sigma', i \sigma } ({\cal H},-\eps) \right)^* 
\nonumber \\
&=&   
\sigma \sigma' {\cal S}^{n m}_{ j \bar{\sigma}', i \bar{\sigma} } (-{\cal H},\eps) \, \nonumber \\
&=&\, m n \, 
\sigma \sigma' \left({\cal S}^{m n}_{ i \bar{\sigma}, j \bar{\sigma}' } (-{\cal H},-\eps) \right)^*, \qquad
\end{eqnarray}
where we introduced sublattice indices $m,n=A (+1), B (-1)$. \\[5mm]

\begin{widetext}

\section{Scattering approach to transport and formulation of the problem}\label{sec:scat}

We consider
a multiterminal device connected to  $i,j=1,2,\ldots N$ electrodes.
The linear response relation is
\begin{eqnarray}\label{eq:slin}
\left(
\begin{array}{c}
J_i \\
I_i^{(0)} \\
I_i^{(x)} \\
I_i^{(y)} \\
I_i^{(z)}
\end{array} \right)
&=&\sum_j
\left(
\begin{array}{ccccc}
\Xi_{ij}^{(00)} & \Gamma_{ij}^{(00)} & \Gamma_{ij}^{(0x)} & \Gamma_{ij}^{(0y)} & \Gamma_{ij}^{(0z)} \\
B_{ij}^{(00)} & G_{ij}^{(00)} & G_{ij}^{(0x)} & G_{ij}^{(0y)} & G_{ij}^{(0z)} \\
B_{ij}^{(x0)} & G_{ij}^{(x0)} & G_{ij}^{(xx)} & G_{ij}^{(xy)} & G_{ij}^{(xz)} \\
B_{ij}^{(y0)} & G_{ij}^{(y0)} & G_{ij}^{(yx)} & G_{ij}^{(yy)} & G_{ij}^{(yz)} \\
B_{ij}^{(z0)} & G_{ij}^{(z0)} & G_{ij}^{(zx)} & G_{ij}^{(zy)} & G_{ij}^{(zz)} 
\end{array}\right)
\left(
\begin{array}{c}
T_j-T_0 \\
V_j-V_0 \\
\mu_j^{(x)}/e \\
\mu_j^{(y)}/e \\
\mu_j^{(z)}/e \\
\end{array}
\right)  \, ,
\end{eqnarray}
between heat, $J_i$, electric, $I_i^{(0)}$ and spin, $I_i^{(\alpha)}$ currents on the one hand, and 
temperatures, $T_j$, voltages, $V_j$ and spin accumulations, $\mu_{j}^{(\alpha)}$ on the other.
\end{widetext}

The coefficients with superindices $^{(00)}$ are the usual thermoelectric coefficients, while the 
coefficients $G^{(\alpha \beta)}$ are conductances and spin-dependent conductances, relating
electric and spin currents to electric voltages and spin accumulations~\cite{ada09,Bar07}. Finally, one has
spin-Peltier matrix elements $\Gamma^{(0\beta)}$ connecting heat currents to spin accumulations 
and spin-Seebeck matrix elements $B^{(\alpha 0)}$ connecting spin currents to temperature differences~\cite{footnote:spinmatrel}.
The resulting spin caloritronic  (spin-Seebeck and spin-Peltier) effects have been investigated 
theoretically~\cite{Dubi,swi,qi,wang} and experimentally~\cite{uchida,Slachter,Flipse}. As usual, 
we assume that there is no spin relaxation in the terminals where
spin currents are measured, so that the latter 
are well defined.
Our goal is to determine reciprocity relations between the elements 
of the Onsager matrix defined on the right-hand side of Eq.~(\ref{eq:slin})
in the ten symmetry classes discussed in Section~\ref{sec:10fold}~\cite{dys62,ver93,alt97}. 

We will express the matrix elements of the Onsager matrix in Eq.~(\ref{eq:slin}) 
in terms of the ${\cal S}$-matrix. We discuss separately purely metallic systems and hybrid systems
consisting of normal metallic components connected to superconductors.

\subsection{Purely metallic systems}

Purely metallic systems fall in either one of the Wigner-Dyson or in one of the chiral
classes.  We start from 
the expression for electric current in
Ref.~[\onlinecite{But86}], extending it to account for heat and spin currents, 
e.g.~along the lines of Refs.~[\onlinecite{But90,ada09}]. 
This gives us the following 
linear relations between electric, heat 
and spin currents, on one hand, and voltages, temperatures 
and spin accumulations, on the other hand;
\begin{subequations}\label{scatt0}
\begin{eqnarray}
J_i &=& \frac{1}{h} \int_{-\infty}^{\infty} {\rm d} \epsilon  \left(-\frac{\partial f}{\partial \epsilon} \right)  
\epsilon \,
\sum_{j, \beta}[2 N_i \delta_{0  \beta} \delta_{ij} -\mathcal{T}^{(0 \beta)}_{ij}(\epsilon)] 
\qquad
\nonumber \\
& & \qquad \qquad \qquad \times 
 [\mu_{j}^{(\beta)} +\delta_{0\beta} \, \epsilon (T_j-T_0)/T_0 ] \, ,
\end{eqnarray}
\begin{eqnarray}
I_i^{(\alpha)} &=& \frac{e}{h} \int_{-\infty}^{\infty} {\rm d} \epsilon \left(-\frac{\partial f}{\partial \epsilon} \right)
\sum_{j, \beta} [2 N_i \delta_{\alpha\beta} \delta_{ij} -\mathcal{T}^{(\alpha\beta)}_{ij}(\epsilon)]
\qquad \nonumber \\
& & \qquad \qquad \qquad \times 
 [\mu_{j}^{(\beta)} +\delta_{0\beta} \, \epsilon (T_j-T_0)/T_0 ] \, , 
\end{eqnarray}
\end{subequations}
where the sums run over all terminal indices $i,j$ and all charge-spin indices $\alpha,\beta=0,x,y,z$.
The electrochemical potential in terminal $j$ is 
$\mu_j^{(0)} = \mu_{\rm F}+ e V_j$  with the applied voltage 
$V_j$ and $T_0$ is the base temperature  about which
the Fermi function $f=(\exp[\epsilon/T]+1)^{-1}$ is expanded. The spin accumulations
$\mu_j^{(\beta)}$, $\beta \neq 0$, are one half times the $\beta$-components of the spin
accumulation vector $\boldsymbol{\mu}_{j}$, giving the difference in chemical
potential between the two spin species along the $\beta$ axis, e.g.
$\mu_j^{(z)} = (\mu_j^{(\uparrow)}-\mu_j^{(\downarrow)})/2$. They are nonequilibrium
spin accumulations whose origin is of little importance here.

In Eqs.~(\ref{scatt0}), we introduced the 
spin-dependent transmission and reflection coefficients
\begin{equation}\label{eq:stran}
\mathcal{T}_{ij}^{(\alpha\beta)} = 
{\rm Tr} [ ({\cal S}_{ij})^\dagger
\sigma^{(\alpha)}_i  {\cal S}_{ij} \  \sigma^{(\beta)}_j],
\end{equation}
where $\sigma^{(\alpha)}$, $\alpha = 0,x,y,z$ are Pauli matrices
($\sigma^{(0)}$ is the identity matrix) and
the trace is taken over both spin and transmission channel indices.
Note the position of the Pauli matrices, where
$\sigma^{(\alpha)}_i$ measures the spin in direction $\alpha$ as the electron exits the systems, while
$\sigma^{(\beta)}_j$ measures it along $\beta$ as the electron enters the system~\cite{Bar07,ada09}.
These coefficients depend on the energy $\epsilon$ of the injected electrons, which we explicitly
wrote in Eq.~(\ref{scatt0}).
Reciprocity relations for the Onsager matrix elements in purely metallic systems
directly follow from combining Eqs.~(\ref{scatt0}) with the
transformation rules for the $\mathcal{T}_{ij}^{(\alpha\beta)}$ under microreversibility.
Pauli matrices satisfy
$\sigma^{(\alpha)}_{\eta \eta'} 
= (-1)^{n_\alpha} \eta \eta' \big[\sigma^{(\alpha)}_{\bar\eta \bar\eta'}\big]^*$
with $n_{x,y,z}=1$ and $n_0=0$. 
Using this and Eq.~(\ref{eq:smatr_lms}) we obtain 
\begin{equation}\label{eq:rules}
\mathcal{T}_{ij}^{(\alpha\beta)}({\cal H},\eps) = (-1)^{n_\alpha+n_\beta} \,
\mathcal{T}_{ji}^{(\beta\alpha)}(-{\cal H},\eps) \, .
\end{equation}
Thus the reciprocity relation
between spin-dependent transmission
coefficients  in Eq.~(\ref{eq:rules}) picks up a minus sign if the spin is resolved upon entering the system,
and another if it is resolved upon leaving the system.

\subsection{Metallic systems with chiral symmetry}

The chiral classes correspond to systems with a bipartite lattice, 
however currently no experiments are capable of measuring sublattice-resolved currents.
Thus the charge and spin-transport is given by Eqs.~(\ref{scatt0},\ref{eq:stran}), with
the trace over channels supplemented by a trace over
the sublattice indices  (A and B sites). 
If one could measure sublattice isospin current, then one would have to
add further Pauli matrices acting in sublattice space into Eqs.~(\ref{eq:stran}), 
leading to extra factors $(-1)^{n_{\alpha'}+n_{\beta'}}$ due to isospin in  Eq.~(\ref{eq:rules}).
We do not consider this possibility further, due to its lack of physical implementation.

A relevant consequence of SLS is however that
from Eq.~(\ref{eq:smatr_chiral_rec}), we get
${\mathcal T}_{ij}^{(\alpha\beta)}(\eps) =  {\mathcal T}_{ji}^{(\beta\alpha)}(-\eps)$. 
Combining this with Eq.~(\ref{eq:rules})  we obtain
\begin{eqnarray}\label{eq:rules-SLS}
\mathcal{T}_{ij}^{(\alpha\beta)}({\cal H},\eps) \ &=& \  
\mathcal{T}_{ji}^{(\beta\alpha)}({\cal H},-\eps) 
\nonumber \\
& = & \ 
(-1)^{n_\alpha+n_\beta} \mathcal{T}_{ji}^{(\beta\alpha)}(-{\cal H},\eps) 
\nonumber \\
& = & \  
(-1)^{n_\alpha+n_\beta} \mathcal{T}_{ij}^{(\alpha\beta)}(-{\cal H},-\eps) \, .
\end{eqnarray}
These relations are strictly valid only insofar as leads are preserving SLS, meaning that
they connect equally to both sublattice sites of each unit cell.

\subsection{Hybrid superconducting-normal metallic systems} 

Hybrid normal-metallic/superconducting systems have Andreev electron-hole scattering. 
This scattering may induce PHS, in which case 
the system falls in one of the four Altland-Zirnbauer symmetry classes
in Table~\ref{table1}~\cite{alt97}.

To include Andreev scattering, one has to consider two kinds of quasiparticles
(electrons and holes), which carry excitation energy
$\pm \epsilon$ counted from the chemical potential of the superconductor 
$\mu_{\rm sc}$.
These quasiparticles are converted into one another when they hit the superconductor.
Ref.~[\onlinecite{Cla96}] constructed a scattering theory of thermoelectric transport which include these effects. 
We need to include spin currents and accumulations.

To do this, we go back to the derivation of the scattering theory
in terms of creation and annihilation
operators acting on scattering states in the lead (see e.g. Ref.~[\onlinecite{Beenakker-review}]).
We write hole creation operators at energy $\veps$,
in terms of electron annihilation operators at energy $-\veps$ as 
$\boldsymbol{c}^{\rm (h)in/out \dagger}_{i;n}(\veps) =
 \boldsymbol{c}^{\rm (e)in/out}_{i;n}(-\veps)$
with
\begin{eqnarray}
\boldsymbol{c}^{\rm (h)in/out \dagger}_{i;n} = 
\left( \! \begin{array}{c}
c^{\rm (h)in/out \dagger}_{i;n \up} \\
c^{\rm (h)in/out \dagger}_{i;n \dn}
\end{array} \! \right) \!, 
 \ \ 
\boldsymbol{c}^{\rm (e)in/out}_{i;n} = 
\left(\! \begin{array}{c}
c^{\rm (e)in/out}_{i;n \up} \\
c^{\rm (e)in/out}_{i;n \dn}
\end{array}\! \right) \, .
\nonumber
\end{eqnarray}
Here, $i$ gives the index of a transverse mode
in the $n$th lead,
while ``in'' and ``out'' indicate whether the wave in that mode is ingoing or outgoing.
As these operators obey fermionic commutation relations,
one has 
\begin{eqnarray}
& & \hskip -15mm
\boldsymbol{c}^{ {\rm e; in} \dagger}_{i;n}(-\veps) \, \boldsymbol{\sigma}^{(\alpha)} \,
\boldsymbol{c}^{  {\rm e; in} }_{i;n} (-\veps) 
\nonumber \\
&=&  
\boldsymbol{\sigma}_0\delta_{\alpha 0} -
\boldsymbol{c}^{ {\rm h; in} \dagger}_{i;n}(\veps) \, 
\sigma^{\rm (\alpha) \rm T} \,
\boldsymbol{c}^{  {\rm h; in} }_{i;n} (\veps) \, ,
\label{Eq:e-to-h}
\end{eqnarray} 
with a similar relation for outgoing  waves.
The transpose in the second term is due to the fact that we commuted the hole operators to 
ensure normal ordering.  We then
use the scattering matrix to write outgoing operators in terms of incoming ones.
Contributions coming from the first term in Eq.~(\ref{Eq:e-to-h}) cancel each other.
We find that the operator which gives the spin-current along axis $\alpha$ in the electron
sector
is $\sigma_\alpha$ [as in Eq.~(\ref{eq:stran})], while it is 
$-\sigma^{\rm T}_\alpha$ in the hole sector. Recalling that 
we use the convention in 
Eqs.~(\ref{eq:bdg},\ref{Eq:Psi-for-e-and-h}), we must also rotate the spin-current operator in the hole
sector.
It becomes $-\sigma^{(y)} \sigma^{(\alpha)\rm T}\sigma^{(y)} = (-1)^{n_\alpha+1} 
\sigma^{(\alpha)}$. 
Thus in this convention, 
we can write this spin-current operator compactly as
$\mu^{n_\alpha+1}\sigma^{(y)}$ which works for both electrons ($\mu=1$) and holes
($\mu=-1$). From here on,
quasiparticle indices $\mu, \nu= +1(e),  -1(h)$ when they appear as prefactors.

This calculation in terms of creation and annihilation operators for electrons and holes
gives us the scattering matrix formula that we desire.
Assuming that the number of transport channels $N_i$ is the same for each quasiparticle
species, Eqs.~(\ref{scatt0}) is replaced by
\begin{widetext}
\begin{subequations}\label{scatt0sc}
\begin{eqnarray}
\label{scatt0scb}
J_i &=& \frac{1}{h} \int_{0}^{\infty} {\rm d} \epsilon  \left(-\frac{\partial f}{\partial \epsilon} \right)  
\epsilon  \, \sum_{j, \beta}  \left\{
[4 N_i  \delta_{ij}-\sum_{\mu,\nu}
\mathcal{T}^{(\mu \nu ; 0 0)}_{ij}(\epsilon)] \, \delta_{0\beta} \, \epsilon (T_j-T_0)/T_0 
 -\sum_{\mu,\nu} \nu \,
\mathcal{T}^{(\mu \nu ; 0 \beta)}_{ij}(\epsilon) \mu_{j}^{(\beta)} 
\right\}
\, , \\ \label{scatt0sca}
I_i^{(\alpha)} &=& \frac{e}{h} \int_{0}^{\infty} {\rm d} \epsilon \left(-\frac{\partial f}{\partial \epsilon} \right)
\sum_{j, \beta} \left\{ 
[4 N_i \delta_{\alpha\beta} \delta_{ij} -\sum_{\mu,\nu} \mu \nu \,
\mathcal{T}^{(\mu \nu ; \alpha\beta)}_{ij}(\epsilon)] \mu_{j}^{(\beta)} 
-\sum_{\mu,\nu} \mu \,
\mathcal{T}^{(\mu \nu ; \alpha 0)}_{ij}(\epsilon) \, \delta_{0\beta} \,  \epsilon (T_j-T_0)/T_0 \right\}
\, , \qquad
\end{eqnarray}
\end{subequations}
\end{widetext}
where the integrals now go over a  range of positive excitation energies~\cite{Cla96} and
we defined $\mu_j^{(0)} = e (V_j-V_{\rm sc})$, i.e. voltages are
measured from the superconducting voltage $V_{\rm sc} = \mu_{\rm sc}/e$. 
We also introduced the spin-dependent, quasi-particle resolved transmission coefficients
\begin{eqnarray}\label{eq:stran_sc}
\mathcal{T}_{ij}^{(\mu \nu; \alpha\beta)} & = &  \
\mu^{n_{\alpha}}\,\nu^{n_\beta}\ 
{\rm Tr} \left[ 
({\cal S}_{ij}^{\mu\nu})^\dagger \sigma^{(\alpha)}_i  
{\cal S}_{ij}^{\mu\nu} \sigma^{(\beta )}_j  \right]
\end{eqnarray}  
where  
${\cal S}_{ij}^{\mu \nu}$ is the block of the ${\cal S}$-matrix corresponding to the transmission
of a quasiparticle of type $\nu=e,h$ 
in lead $j$ to a $\mu$-quasiparticle in lead $i$.

\begin{table*}
\begin{tabular}{| l  |  l ||  c  | } 
\hline
\multicolumn{2}{|c||}{{\bf Symmetry class}}   & {\bf Seebeck-Peltier Onsager relations} \\ 
\multicolumn{2}{|c||}{}  &\ \  {\bf  from microreversibility} \ \ \\ 
\hline \hline
Wigner-Dyson & \ A (unitary) & $B_{ij}^{(\beta 0)}({\cal H}) T_0 = (-1)^{n_\beta} \Gamma_{ji}^{(0\beta)} (-{\cal H})$ \\
 & \ AI (orthog.) & $B_{ij}^{(\beta 0)} T_0 = \Gamma_{ji}^{(0\beta)}  
\propto \delta_{0 \beta }  $
 \\
& \ AII (sympl.) & $B_{ij}^{(\beta 0)} T_0 = (-1)^{n_\beta} \Gamma_{ji}^{(0\beta)}$ \\[1mm]
\hline \hline
Chiral & \ AIII(unitary) & $B_{ij}^{(\beta 0)}({\cal H}) T_0 = (-1)^{n_\beta} \Gamma_{ji}^{(0\beta)} (-{\cal H})$ \\
\  & \ BDI (orthog.) & $B_{ij}^{(\beta 0)} T_0 = \Gamma_{ji}^{(0\beta)}   
 \propto \delta_{0 \beta }  $
\\
\ & \ CII (sympl.) &  $B_{ij}^{(\beta 0)} T_0 = (-1)^{n_\beta} \Gamma_{ji}^{(0\beta)}$ \\[1mm]
\hline \hline
Altland-Zirnbauer \  
&\ D&  $B_{ij}^{(\beta 0)}({\cal H}) T_0 = (-1)^{n_\beta} \Gamma_{ji}^{(0\beta)} (-{\cal H})$ \\
\ 
&\ C    &  $B_{ij}^{(\beta 0)}({\cal H}) T_0 = \Gamma_{ji}^{(0\beta)} (-{\cal H})$ \\
&\ DIII &  $B_{ij}^{(\beta 0)} T_0 = (-1)^{n_\beta} \Gamma_{ji}^{(0\beta)}$\\
&\ CI   & $B_{ij}^{(\beta 0)} T_0 = \Gamma_{ji}^{(0\beta)} 
 \propto \delta_{0 \beta } $ 
\\[1mm]
\hline 
\end{tabular}
\caption{\label{table3}
The Onsager reciprocity relations
arising from microreversibility and which involve
the Peltier and spin-Peltier matrix elements, $\Gamma^{(0\beta)}$,
and the Seebeck and spin-Seebeck matrix elements, $B^{(\beta 0)}$.
These relations, combined with those due to PHS or SLS in Table~\ref{table5}, give the
complete set of Onsager relations for each symmetry class.}
\end{table*}

The main novelty brought about by superconductivity is that the elements
of the Onsager matrix now depend on Andreev processes via hybrid
transmission coefficients $\mathcal{T}_{ij}^{(eh; \alpha\beta)}$
and $\mathcal{T}_{ij}^{(he; \alpha\beta)}$, which
contribute differently to heat versus electric and spin currents --- see in particular
the last terms in Eqs.~(\ref{scatt0scb}) and (\ref{scatt0sca}).
From Eq.~(\ref{eq:smatr_lms}) one obtains
\begin{equation}\label{eq:rules_sc}
\mathcal{T}_{ij}^{(\mu\nu;\alpha\beta)}({\cal H},\eps) = (-1)^{n_\alpha+n_\beta} 
\mathcal{T}_{ji}^{(\nu \mu;\beta\alpha)}(-{\cal H},\eps) \, ,
\end{equation}
which extends Eq.~(\ref{eq:rules}) to include superconductivity.

Eq.~(\ref{eq:rules_sc}) applies to any hybrid system, regardless of whether PHS
is present or not. If additionally, the system has unbroken PHS,
then the scattering matrix obeys 
Eq.~(\ref{eq:smatr_sc}), i.e. 
${\cal S}^{\mu\nu}({\cal H}) = \mu \nu \,\sigma^{(y)} [{\cal S}^{\bar{\mu} \bar{\nu}}({\cal H})]^*
\sigma^{(y)} $,
where, as before, $\mu, \nu= +1(e),  -1(h)$.
(Ref.~[\onlinecite{Cla96}] has this formula for SRS, where ${\cal S}^*$ commutes with  $\sigma^{(y)}$).
We substitute this into Eq.~(\ref{eq:rules_sc}), and then
substitute $\sigma^{(y)}\sigma^{(\alpha)}\sigma^{(y)}= (-1)^{n_\alpha} \sigma^{(\alpha) \rm T}$.
Observing that the trace is invariant under the transpose of its argument,
we find that PHS gives
\begin{eqnarray}\label{eq:rules_phs}
\mathcal{T}_{ij}^{(\mu\nu;\alpha\beta)}({\cal H},\eps) 
\ &=&\ \mathcal{T}_{ij}^{(\bar{\mu}\bar{\nu};\alpha\beta)}({\cal H},\eps)
\nonumber \\
&=&\  (-1)^{n_\alpha+n_\beta} \,\mathcal{T}_{ji}^{(\nu \mu;\beta\alpha)}(-{\cal H},\eps)
\nonumber \\
&=&\  (-1)^{n_\alpha+n_\beta}  \,\mathcal{T}_{ji}^{(\bar{\nu} \bar{\mu};\beta\alpha)}(-{\cal H},\eps) \, .
\qquad
\end{eqnarray}

\section{Onsager relations}\label{sec:onsager} 

Eqs.~(\ref{scatt0}), (\ref{eq:rules}), (\ref{scatt0sc}), (\ref{eq:rules_sc}) and (\ref{eq:rules_phs})
are all we need to derive reciprocity relations
between the coefficients of the Onsager matrix in Eq.~(\ref{eq:slin}). 
Tables~\ref{table3}, \ref{table4} and \ref{table5} provide a complete list of all Onsager reciprocity
relations for coupled electric, thermoelectric and spin transport
in single-particle Hamiltonian systems.
The Onsager relations which can be derived from microreversibility
are divided into two sets.
Firstly, Table~\ref{table3} gives the Peltier/Seebeck relations,
between coefficients $\Gamma^{(0 \beta)}$ and $B^{(\beta 0)}$. Secondly
Table~\ref{table4} gives the reciprocity relations for conductances 
$G_{ij}^{(\alpha \beta)}$, $\Xi_{ij}^{(\alpha \beta)}$.
As an example,  we note that for both the Wigner-Dyson and chiral orthogonal classes, 
the presence of SRS imposes $\mathcal{T}_{ij}^{(\alpha \beta)}=
\mathcal{T}_{ij}^{(0 0)} \delta_{\alpha \beta}$, while TRS gives $\mathcal{T}_{ij}^{(\alpha \beta)} =
\mathcal{T}_{ji}^{(\beta \alpha)}$. Therefore, when both symmetries are present in those classes,
$X_{ij}^{(\alpha \beta)} = X_{ji}^{(\alpha \beta)}$, for $X=\Xi$, $\Gamma$, $B$ and $G$.
In addition there are those Onsager relations which can be derived from
either the conservation of quasiparticle species (absence of Andreev processes turning e into h, and vice-versa), 
or from the presence of PHS or SLS. They are listed in Table~\ref{table5}.

\begin{table*}
\begin{tabular}{| l  |  l ||  c  | } 
\hline
\multicolumn{2}{|c||}{{\bf Symmetry class}}   & \ \ {\bf Onsager relations between conductances, $X=G, \Xi$} \ \ \\ 
\multicolumn{2}{|c||}{}  & {\bf  from microreversibility} \\ 
\hline \hline
Wigner-Dyson & \ A (unitary) & $X_{ij}^{(\alpha \beta)}({\cal H}) = (-1)^{n_\alpha+n_\beta} X_{ji}^{(\beta \alpha)}(-{\cal H})$ \\
 & \ AI (orthog.) & $X_{ij}^{(\alpha \beta)} =  X_{ji}^{(\beta \alpha)} 
\propto  \delta_{\al\be}   $
\\
& \ AII (sympl.) & $X_{ij}^{(\alpha \beta)} = (-1)^{n_\alpha+n_\beta} X_{ji}^{(\beta \alpha)}$ \\[1mm]
\hline \hline
Chiral & \ A III(unitary) & $X_{ij}^{(\alpha \beta)}({\cal H}) = (-1)^{n_\alpha+n_\beta} X_{ji}^{(\beta \alpha)}(-{\cal H})$ \\
\  & \ BDI (orthog.) & $X_{ij}^{(\alpha \beta)} =  X_{ji}^{(\beta \alpha)}
\propto \delta_{\al\be} $
 \\
\ & \ CII (sympl.) &  $X_{ij}^{(\alpha \beta)} = (-1)^{n_\alpha+n_\beta} X_{ji}^{(\beta \alpha)}$ \\[1mm]
\hline \hline
Altland-Zirnbauer \  
&\ D&  $X_{ij}^{(\alpha \beta)}({\cal H}) = (-1)^{n_\alpha+n_\beta} X_{ji}^{(\beta \alpha)}(-{\cal H})$ \\ 
&\ C    &  $X_{ij}^{(\alpha \beta)}({\cal H}) = X_{ji}^{(\beta \alpha)}(-{\cal H})$ \\
&\ DIII &  $X_{ij}^{(\alpha \beta)} = (-1)^{n_\alpha+n_\beta} X_{ji}^{(\beta \alpha)}$\\
&\ CI   & $X_{ij}^{(\alpha \beta)} =  X_{ji}^{(\beta \alpha)} 
\propto \delta_{\al\be}  $ 
 \\[1mm]
\hline 
\end{tabular}
\caption{\label{table4}
The Onsager reciprocity relations
arising from microreversibility and which involve
the electrical and spin-dependent
conductances $G^{(\alpha \beta)}$ and the heat conductance $\Xi^{(00)}$,
These relations, combined with those due to PHS or SLS in Table~\ref{table5}, give  the
complete set of Onsager relations for each symmetry class.
}
\end{table*}

Some important features are that
(i) in multiterminal devices one needs to consider conductance, Seebeck and Peltier matrices, and
the reciprocity relations require to take their transpose, the latter operation being tantamount
to momentum inversion as required by microreversibility,
(ii) spin transport introduces additional minus signs everytime a spin is measured,
(iii) exact PHS leads to the disappearance of thermoelectric and spin caloritronic effects,
(iv) at half-filling, 
exact SLS  leads to the disappearance of thermoelectric but {\it not} spin caloritronic effects.

That thermoelectric and spin caloritronic
effects vanish in the presence of PHS directly follows from
Eq.~(\ref{eq:stran_sc}) that transmission coefficients satisfy $\mathcal{T}_{ij}^{(\mu \nu; \alpha\beta)}
= \mathcal{T}_{ij}^{(\bar{\mu} \bar{\nu}; \alpha\beta)}$ when PHS is strictly enforced.
This gives in particular $\sum_\nu \nu \mathcal{T}_{ij}^{(\mu \nu ; \alpha\beta)}
= \sum_\mu \mu \mathcal{T}_{ij}^{(\mu \nu ; \alpha\beta)} = 0$ which, together with
Eq.~(\ref{scatt0sc}), directly gives $B_{ij}^{(0 \alpha)}({\cal H}) = 
\Gamma_{ij}^{(\alpha 0)}({\cal H}) = 0$. 

The vanishing of thermoelectric effects with PHS is reminiscent of Mott's 
relation, giving that the Seebeck coefficient is proportional to  
the derivative of the conductance at the Fermi energy 
-- the latter vanishes in PHS systems. Still, hybrid normal metallic/superconducting 
systems often exhibit larger thermoelectric
effects than their purely metallic counterpart, which typically happens in the crossover
regime between Altland-Zirnbauer and Wigner-Dyson symmetry classes.
For the crossover systems described in Table~\ref{table2},
thermoelectric effects can be quite large~\cite{Eom98,Cad09,Par03}. 

We close this section with two comments on SLS at half-filling, when the chemical potential is
at zero energy.   Systems in the chiral symmetry classes
have transmission coefficients with extra symmetries given in Eq.~(\ref{eq:rules-SLS}).
The latter have important consequences for the symmetry of transport,
if the trace over the sublattice index in Eq.~(\ref{eq:stran}) 
involves only pairs of sublattice sites, i.e. 
when SLS is not broken by the terminals. When this is the case,
the first and second equalities in Eq.~(\ref{eq:rules-SLS}), together with Eqs.~(\ref{scatt0}), 
give
$G_{ij}^{(\alpha \beta)}({\cal H})= G_{ji}^{(\beta \alpha)}({\cal H})$ 
and $G_{ij}^{\alpha \beta}({\cal H}) = (-1)^{n_\alpha+n_\beta} G_{ji}^{\beta\alpha}(-{\cal H})$
respectively, where we recall that $n_0=0$ and $n_{x,y,z}=1$.
We obtain identical results for $\Xi_{ij}^{(00)}$, and thus conclude that
 \begin{eqnarray} \label{eq:1}
& & G_{ij}^{(\alpha \beta)}({\cal H})  = (-1)^{n_\alpha+n_\beta} 
G_{ij}^{(\alpha \beta)}(-{\cal H}), 
\nonumber \\
& & 
\Xi_{ij}^{(00)}({\cal H})  = \Xi_{ij}^{(00)}(-{\cal H}) \, .
\end{eqnarray}
We see that charge-conductance, spin-conductances and thermal conductance 
are even in external fields, ${\cal H}$, 
irrespective of how many terminals the device has. This is in contrast to
normal-metallic systems without SLS, where 
only two-terminal devices have conductances even in ${\cal H}$. 
In contrast, spin-to-charge and charge-to-spin conversion are strictly odd in ${\cal H}$,
irrespective of how many terminals the device has. 

\begin{table*}[t]
\begin{tabular}{| l  |  l ||  c  | } 
\hline
\multicolumn{2}{|c||}{{\bf Symmetry class}}   & {\bf Special additional relations } \\ 
\hline \hline
Wigner & \ A (unitary) & $B_{ij}^{(00)}({\cal H}) = B_{ji}^{(00)}(-{\cal H})$ \\
\ -Dyson & \ AI (orthog.) & $B_{ij}^{(00)}  = B_{ji}^{(00)} $ \\
& \ AII (sympl.) & $B_{ij}^{(00)}  = B_{ji}^{(00)} $ \\[1mm]
\hline \hline
Chiral  & \ AIII(unitary) & 
\ \ $X_{ij}^{(\alpha\beta)}({\cal H})= X_{ij}^{(\alpha\beta)}(-{\cal H})$ for $X =\{G,\Xi\}$,\ \ 
$B_{ij}^{(\beta 0)}({\cal H})= B_{ij}^{(\beta 0)}(-{\cal H})$, \ \ $B_{ij}^{(0 0)}({\cal H})= 0$\ \ \\
\ (half-filling only) \  \  & \ BDI (orthog.) & \ \ $B_{ij}^{(\beta 0)}=\Gamma_{ij}^{(0 \beta)} = 0$ \ \ \\
\ & \ CII (sympl.) & $B_{ij}^{(\beta 0)} = \Gamma_{ij}^{(0 \beta)} = 0$   \\[1mm]
\hline \hline
Altland 
&\ D&  $B_{ij}^{(\beta 0)}({\cal H})=\Gamma_{ij}^{(0 \beta)}({\cal H}) = 0$\,  \\
\ -Zirnbauer 
&\ C    &  $B_{ij}^{(\beta 0)}({\cal H}) =\Gamma_{ij}^{(0 \beta)}({\cal H}) = 0$  \\
&\ DIII &  $B_{ij}^{(\beta 0)}=\Gamma_{ij}^{(0 \beta)} = 0$ \\
&\ CI   & $B_{ij}^{(\beta 0)} = \Gamma_{ij}^{(0 \beta)} = 0$  \\[1mm]
\hline 
\end{tabular}
\caption{\label{table5}
Additional reciprocity relations induced by the conservation of each quasiparticle species 
(absence of Andreev reflection from e to h), 
or by the presence of PHS or SLS,
to be added to the Onsager relations in Tables~\ref{table3} and \ref{table4}.
The additional relation between $B$ and its transpose (and an identical one, not listed here, 
between $\Gamma$ and its transpose)
in the Wigner-Dyson classes
was first noticed in Ref.~[\onlinecite{But90}], which allows one to express the Seebeck/Peltier relation of
Table~\ref{table3} in two different but equivalent ways. 
Thermoelectric and spin caloritronic effects
disappear identically in the presence of PHS.
The relations in the chiral classes correspond to transport at half-filling, $E=0$.
Outside of this regime, chiral systems have the same relations as in the (corresponding) 
Wigner-Dyson classes.}
\end{table*}

Turning to thermoelectric and spin caloritronic effects,
the first equality in Eq.~(\ref{eq:rules-SLS}) gives
$B_{ij}^{(\beta 0)}({\cal H})\,T_0 = -\Gamma_{ji}^{(0\beta)}({\cal H})$, 
while the second equality  in Eq.~(\ref{eq:rules-SLS}) gives us the usual relation
$B_{ij}^{(\beta 0)}({\cal H})\,T_0 = (-1)^{n_\beta} \Gamma_{ji}^{(0\beta)}(-{\cal H})$.
Thus we can conclude that
\begin{subequations}\label{eq:2}
 \begin{eqnarray} 
& & B_{ij}^{(\beta 0)}({\cal H})  = B_{ij}^{(\beta 0)}(-{\cal H})  \ \hbox{ for }\ \beta \in \{x,y,z\}, 
\\
& & 
B_{ij}^{(00)}({\cal H}) = 0, 
\end{eqnarray}
\end{subequations}
with identical relations for $\Gamma_{ij}^{(0\beta)}$.
Additionally, any system with PHS
has no thermoelectric nor spin caloritronic response.
Looking at Table~\ref{table1} we see that only the AIII symmetry class has  SLS without PHS.
Thus in this symmetry class, the spin-Seebeck and spin-Peltier coefficients are 
even functions of the external field, ${\cal H}$, while 
the usual Seebeck and Peltier coefficients vanish identically.

We stress, however, that the analysis leading to 
Eqs.~(\ref{eq:1}) and (\ref{eq:2}) holds only at half-filling, when the Fermi function in 
Eq.~(\ref{scatt0}) is symmetric around $\eps=0$, and, perhaps physically more important,
when the terminals do not break SLS. This requires leads to be connected with equal strength
to both sublattice sites in each unit cell.

\section{Examples of reciprocity relations in spintronics and spin caloritronics}
\label{sec:examples-spin}

\subsection{Spin Hall and inverse spin Hall effects}

\begin{figure*}
\includegraphics[width=10.5cm]{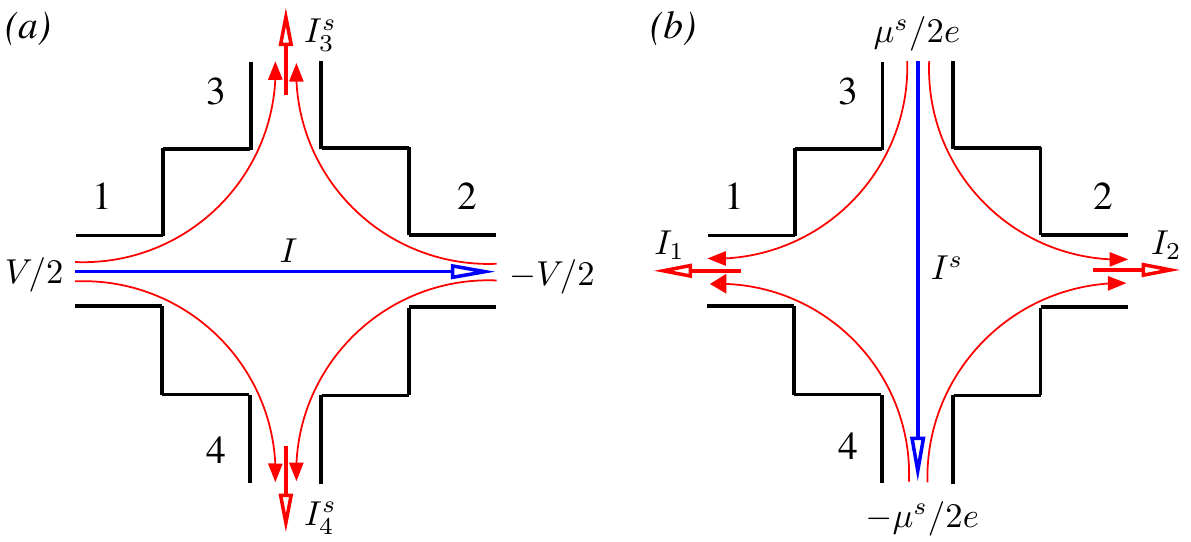}
\caption{\label{fig:she} Sketch of a four-terminal spin-Hall (a) and inverse
spin Hall (b) experiment. (a) In the spin Hall effect, an electric current (blue arrow) 
generates transverse spin currents (red) via the action of spin-orbit coupling. (b) In the
inverse spin Hall effect spin accumulations inject a spin current (blue arrow) which, in the presence of
spin-orbit coupling, generates a transverse electric current (red).}
\end{figure*}

As a first example of the reciprocities we derived, we discuss the spin 
Hall~\cite{han05,Bar07,DP71,eng07,Kato}
and the inverse spin Hall~\cite{han05,val06,Sai06,Kim07,Sek08,Liu11,ada09} effects. The two effects are sketched in Fig.~\ref{fig:she}.
In the spin Hall effect, Fig.~\ref{fig:she}a,
one passes an electric current between
terminals 1 and 2 and measures the spin current between terminals 3 and 4. 
The voltages at terminals 3 and 4 are set such that no current flows through them
on time average. In the limit of large and identical number of channels in each terminal, $N \gg 1$, 
the voltages $V_{3}$ and $V_4$ lie almost exactly in the middle between $V_1$ and $V_2$, 
$V_{3,4} \simeq (V_1+V_2)/2$ for ballistic systems~\cite{Bar07}.  
We assume that this is the case here, and set
$V_1=V/2$, $V_2=-V/2$, $V_{3,4}=0$.

The presence of spin-orbit coupling inside the system generates a spin current 
flowing through the transverse terminals. When all terminals are at zero temperature,
these currents are given by
\begin{subequations}
\begin{eqnarray}
I_{3}^{(\alpha)} & = & -\frac{e^2 V}{2h} ( \mathcal{T}_{31}^{(\alpha 0)} - \mathcal{T}_{32}^{(\alpha 0)} ) \, , \\
I_{4}^{(\alpha)} & = & -\frac{e^2 V}{2h} ( \mathcal{T}_{41}^{(\alpha 0)} - \mathcal{T}_{42}^{(\alpha 0)} ) \, .
\end{eqnarray}
\end{subequations}
In the inverse spin Hall effect, Fig.~\ref{fig:she}b, there is no voltage bias, but instead
terminals 3 and 4 have opposite spin accumulations. In an idealized situation they will be  
$\pm \mu^{(\alpha)}/2$.
Spin-orbit coupling converts this spin accumulation into a transverse electric current.
The currents in terminals 1 and 2 read
\begin{subequations}
\begin{eqnarray}
I_{1}^{(0)} & = & -\frac{e \mu^{(\alpha)}}{2h} ( \mathcal{T}_{13}^{(0 \alpha)} - \mathcal{T}_{14}^{(0 \alpha)} ) \, , \\
I_{2}^{(0)} & = & -\frac{e \mu^{(\alpha)}}{2h} ( \mathcal{T}_{23}^{(0 \alpha)} - \mathcal{T}_{24}^{(0 \alpha)} ) \, .
\end{eqnarray}
\end{subequations}
In both cases, the Hall part of the currents, flowing between 3 and 4 in the case of the spin Hall
effect and betwee 1 and 2 in the case of the inverse spin Hall effect, 
is given by the difference in the two currents. We define spin Hall and inverse spin Hall conductances as 
$I_{sHe}^{(\alpha)} = I_{3}^{(\alpha)} -I_{4}^{(\alpha)} = G_{\rm sHe}  V$ and 
$I_{isHe}^{(0)}  =  I_{2}^{(0)} -I_{1}^{(0)} =  G_{\rm isHe} \, \mu^{(\alpha)} /e$.
One obtains
\begin{subequations}
\begin{eqnarray}\label{eq:rec_she}
G_{\rm sHe}  &=& -\frac{e^2}{2h} ( \mathcal{T}_{31}^{(\alpha 0)} - \mathcal{T}_{32}^{(\alpha 0)} - \mathcal{T}_{41}^{(\alpha 0)} +\mathcal{T}_{42}^{(\alpha 0)} )
\, , \qquad \\
G_{\rm isHe}  &=& -\frac{e^2}{2h} (\mathcal{T}_{23}^{(0 \alpha)} - \mathcal{T}_{24}^{(0 \alpha)}
-
\mathcal{T}_{13}^{(0 \alpha)} + \mathcal{T}_{14}^{(0 \alpha)} )
\, .
\end{eqnarray}
\end{subequations}
Together with Eq.~(\ref{eq:rules}), Eq.~(\ref{eq:rec_she}) gives $G_{\rm sHe} = G_{\rm isHe}$.
The reciprocity between direct and inverse spin Hall conductances
is exact and does not require sample averaging, as sometimes claimed~\cite{han05}.

\subsection{Reciprocity between spin injection and magnetoelectric spin currents}

For a spin index $\beta=0$, 
Eqs.~(\ref{eq:rules}) and (\ref{eq:rules_sc}) establish the reciprocity between 
magnetoeletric effects generating spin currents from electric voltage biases and
spin injection from spin accumulations in the terminals, a special case of which is the above-discussed
spin Hall effect/inverse spin Hall effect reciprocity. In the presence of TRS,
it has already been observed that one consequence of Eq.~(\ref{eq:smatr_lms})
is that no spin current can be magnetoelectrically generated in a two-terminal 
device if the exit lead carries a single (spin-degenerate) transport channel. 
The reciprocity relations of Eqs.~(\ref{eq:rules}) and (\ref{eq:rules_sc}) further impose
that a spin injection from such a terminal is incapable of generating an
electric current, unless one goes to the nonlinear regime~\cite{stano2}.
This seems not to have been noted so far. 

\subsection{Spin Seebeck and spin Peltier coefficients in two-terminal geometries}

In two-terminal geometries, the electric conductance is symmetric in TRS breaking fields,
which follows from current conservation or gauge invariance, 
together with the symmetry of electric reflection
coefficients, $G_{ii}^{(00)}({\cal H}) = G_{ii}^{(00)}(-{\cal H})$ (see e.g. Ref.~[\onlinecite{But86}]).
Including spin-transport, the unitarity of the scattering matrix further results in 
spin-current conservation and generalized gauge invariance,  
\begin{eqnarray}\label{Eq:general-gauge-invariance}
& & \sum_i \left(2N_i \delta_{0\alpha}\delta_{ij}-\mathcal{T}_{ij}^{(0 \alpha)} \right)= 0 \, ,
\qquad
\nonumber   \\  
& & \sum_j  \left(2N_i \delta_{0\alpha}\delta_{ij}-\mathcal{T}_{ij}^{(\alpha 0)} \right)= 0 \, ,
\end{eqnarray} 
under the assumption that
the number of transport channels coupling the
system to external reservoirs is spin-independent.
In two-terminal geometries this gives
\begin{eqnarray}\label{eq:conservations}
& & B_{11}^{(\beta 0)} + B_{12}^{(\beta 0)} = B_{21}^{(\beta 0)} + B_{22}^{(\beta 0)} = 0 \, , 
\qquad
\nonumber   \\  
& & \Gamma_{11}^{(0 \beta)} + \Gamma_{21}^{(0 \beta)} = \Gamma_{12}^{(0 \beta)} + 
\Gamma_{22}^{(0 \beta)} = 0 \, .
\end{eqnarray}
However, unlike for the charge conductance, the thermoelectric reflection coefficients
can have both a symmetric and an antisymmetric component. This is directly seen from the 
expression 
\begin{eqnarray}
B_{ii}^{(\beta 0)} ({\cal H}) = \frac{-e}{h} \int {\rm d} \epsilon \left(-\frac{\partial f}{\partial \epsilon} \right)
\, \frac{\epsilon}{T_0} \,  
\mathcal{T}^{(\beta 0)}_{ii}({\cal H},\epsilon) 
\end{eqnarray}
for the spin Seebeck reflection coefficient. For example for
$\beta=z$, the spin-dependent transmission coefficient in the integrand reads
\begin{subequations}
\begin{eqnarray}
{\mathcal T}_{ii}^{(z0)} &=& {\mathcal T}_{\rm s}^{(z0)} - {\mathcal T}_{a}^{(z0)} \, , \\
 {\mathcal T}_{\rm s}^{(z0)} & \equiv & T_{i\uparrow,i\downarrow}-T_{i\downarrow,i\uparrow}  \, \\
 {\mathcal T}_{a}^{(z0)} & \equiv & T_{i\uparrow,i\uparrow}- T_{i\downarrow,i\downarrow}\, , 
\end{eqnarray}
\end{subequations}
which, from Eq.~(\ref{eq:smatr_lms})  
has both symmetric, $ {\mathcal T}_{\rm s}^{(z0)}({\cal H}) = 
{\mathcal T}_{\rm s}^{(z0)}(-{\cal H})$, and antisymmetric,  $ {\mathcal T}_{a}^{(z0)}({\cal H}) =
{\mathcal T}_{a}^{(z0)}(-{\cal H})$ components.

An interesting example is provided by 
a two-terminal system with a well-defined spin quantization axis. 
This is the case, for example, 
for a system without spin-orbit coupling in a uniform 
Zeeman field, for two-dimensional systems with both Rashba and Dresselhaus
spin-orbit interactions of equal strengths~\cite{egues}, or for a system
with pure $\, \vec{l} \cdot \vec{s} \,$ spin-orbit coupling.
Without loss of generality we define the spin quantization axis as the $z$-axis.
Then ${\cal S}$ commutes with $\sigma^{(z)}$, i.e.~it is diagonal in spin space.
From Eq.~(\ref{eq:stran}), we find that ${\mathcal  T}_{ij}^{(z 0)}({\cal H}) = {\mathcal T}_{ij}^{(0 z)}({\cal H})$ 
and ${\mathcal T}_{ij}^{(\alpha 0)}({\cal H}) =0$ when $\alpha=x,y$.
Combining this with Eqs.~(\ref{Eq:general-gauge-invariance}), we have 
${\mathcal T}_{12}^{(z 0)}({\cal H})= {\mathcal T}_{12}^{(0 z)}({\cal H})=
{\mathcal T}_{21}^{(z 0)}({\cal H})= {\mathcal T}_{21}^{(0 z)}({\cal H})$.
Thus 
\begin{eqnarray}
B^{(z 0)}_{12}({\cal H}) T_0= \Gamma^{(0 z)}_{12} ({\cal H}) \, ,
\end{eqnarray}
with $\Gamma^{(0\alpha)}_{\rm 12} ({\cal H})=0$ when $\alpha=x,y$.  
Next we recall that the Seebeck-Peltier
Onsager relations contain an extra minus sign for spin caloritronic effects 
compared to usual thermoelectric effects (see Table~\ref{table3}).
This  extra minus sign means that
$B^{(z0)}_{\rm 12}$ and $\Gamma^{(0z)}_{\rm 12}$ are odd in ${\cal H}$, 
while
$B^{(00)}_{\rm 12}$ and $\Gamma^{(00)}_{\rm 12}$ are even in ${\cal H}$. 
Thus any two-terminal system with a spin-quantization axis will have
spin-Seebeck and spin-Peltier effects which are odd functions of TRS breaking
fields, 
while the normal Seebeck and Peltier effects are even function of those fields.

\section{Examples of reciprocity relations 
in thermoelectricity with hybrid systems}
\label{sec:examples-sc}

Thermoelectric effects in the presence of superconductivity, 
in particular the thermopower $S=-B^{(00)} /G^{(00)} $ and thermal conductance
$\Xi^{(00)}$, have attracted quite some experimental~\cite{Eom98,Cad09,Par03} and theoretical interest~\cite{Cla96,Vir04,Tit08,Jac10,Eng11,Sev00,Bez03}. 
However, the exact form that the Seebeck-Peltier
Onsager reciprocity relation takes has never been clarified, 
despite the fact that  two-terminal devices with superconductors
usually exhibit odd Seebeck coefficients
$S({\cal H}) = - S(-{\cal H})$, 
in stark contrast with Mott's relation~\cite{Ash67}.
Mott's relation between the thermopower of metallic systems at low temperature 
and the energy derivative of the conductance at
the Fermi energy, reads
\begin{equation}
S = -\frac{\pi^2 k_{\rm B}^2 T}{3 e} \partial_E {\rm ln} G(E_{\rm F}) \, ,
\end{equation}
and thereby indicates that $S$ should be even in ${\cal H}$.
This evenness of $S$ is confirmed by the scattering theory for metallic systems \cite{But90}.
In this section we provide examples clarifying this issue
using scattering theory to show that 
 $S$ can have any symmetry under ${\cal H}\to -{\cal H}$
when superconductors are present.

\begin{figure}[t]
\includegraphics[width=5.5cm]{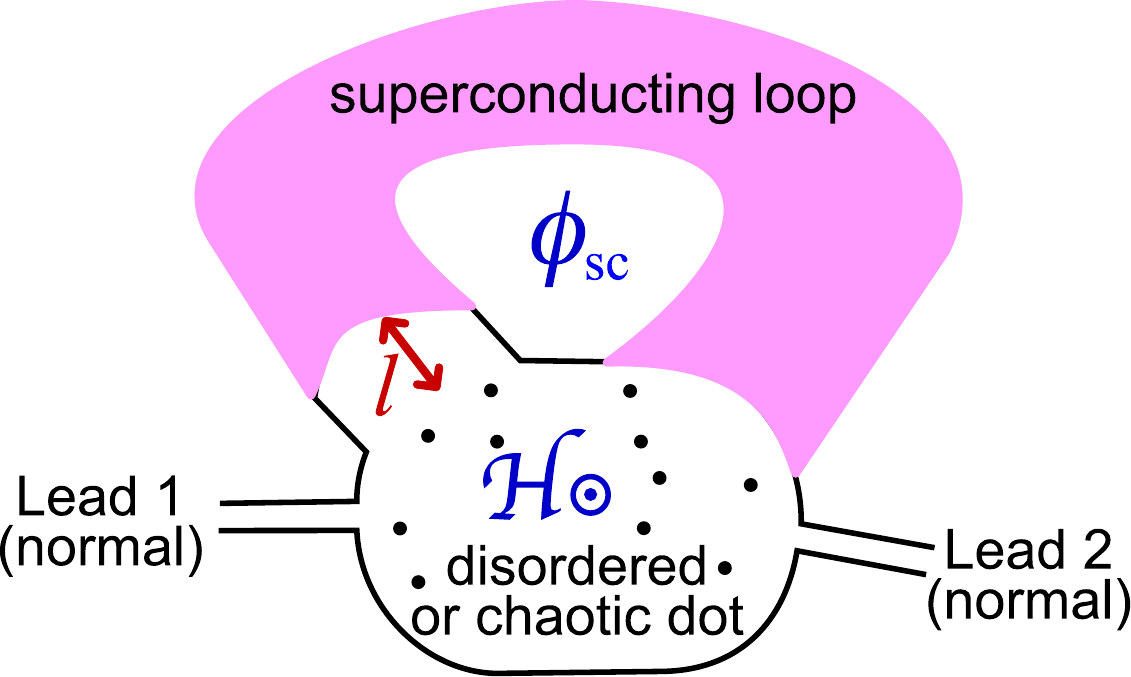}
\caption{\label{Fig:SC}
An Andreev interferometer: two-terminal hybrid 
system with two normal metallic/superconducting contacts with tunable superconducting
phase difference $\phi_{\rm SC}$. The latter, together with a systematic delay (indicated
by the extra length $\ell$) towards one of the superconducting contacts
can break PHS and generate finite thermoelectric effects~\cite{Jac10}.
The particular symmetry class in which the system falls
is given by how much magnetic field ${\cal H}$
and how much spin-orbit coupling there is in the dot, see Table \ref{table1}.}
\end{figure}

\subsection{Seebeck--Peltier reciprocity relation}

Andreev scattering strongly influences the Seebeck--Peltier reciprocity relation between
$\Gamma$ and $B$ coefficients.
Comparison of the last terms
in Eqs.~(\ref{scatt0scb}) and (\ref{scatt0sca}) shows that the $ee$ and $hh$ terms 
in $\Gamma$ and $B$ have the
same sign, while the $eh$ and $he$ terms acquire a relative minus sign. 
This breaks one of
the Onsager relations between Peltier  and Seebeck  coefficients. For metallic
systems, one has both
$B_{ij}^{(0 0)}({\cal H}) T_0 = \Gamma_{ji}^{(0 0)}(-{\cal H})$ and
$B_{ij}^{(0 0)}({\cal H}) T_0 = \Gamma_{ij}^{(0 0)}({\cal H})$~\cite{But90}, 
however, with superconductivity,
only $B_{ij}^{(0 0)}({\cal H}) T_0 = \Gamma_{ji}^{(0 0)}(-{\cal H})$ holds.

When PHS strictly holds, however, $\sum_\nu \nu {\mathcal T}_{ij}^{(\mu \nu; \alpha \beta)} = 
\sum_\mu \mu {\mathcal T}_{ij}^{(\mu \nu; \alpha \beta)} =0$ and both $\Gamma$-
and $B$-coefficients vanish identically, regardless of the temperature. However, interesting thermoelectric
effects appear in hybrid systems when PHS is broken. Focusing on a two-terminal geometry, as depicted
in Fig.~\ref{Fig:SC}, Eq.~(\ref{scatt0sc}) can be rewritten in the form
\begin{eqnarray}
\left(
\begin{array}{c}
J\\
I
\end{array} \right)
&=&
\left(
\begin{array}{cc}
\Xi & \Gamma\\
B & G
\end{array} \right)
\left(
\begin{array}{c}
\Delta T\\
\Delta V
\end{array} \right) \, , 
\end{eqnarray}
which depends only on the voltage and temperature differences between the two normal reservoirs.
The two-terminal thermoelectric coefficients are given by
\begin{subequations}
\begin{eqnarray}
G 
&=& G^{(00)}_{11}
 - {\big( G^{(00)}_{11}+G^{(00)}_{12}\big) \big( G^{(00)}_{11} +G^{(00)}_{21}\big)
\over  G_{11}^{(00)}+G_{22}^{(00)}+G_{12}^{(00)}+G_{21}^{(00)}  } \, , \qquad
\\
\Xi &=& \Xi^{(00)}_{11} 
-{\big( \Gamma^{(00)}_{11}+\Gamma^{(00)}_{12}\big) \big( B^{(00)}_{11} +B^{(00)}_{21}\big)
\over  G_{11}^{(00)}+G_{22}^{(00)}+G_{12}^{(00)}+G_{21}^{(00)}  }\, ,
\\
B 
&=& B^{(00)}_{11}
- {\big( G^{(00)}_{11}+G^{(00)}_{12}\big) \big( B^{(00)}_{11} +B^{(00)}_{21}\big)
\over  G_{11}^{(00)}+G_{22}^{(00)}+G_{12}^{(00)}+G_{21}^{(00)} } \, ,
\\
\Gamma 
&=& \Gamma^{(00)}_{11} 
-{\big( \Gamma^{(00)}_{11}+\Gamma^{(00)}_{12}\big) \big( G^{(00)}_{11} +G^{(00)}_{21}\big)
\over  G_{11}^{(00)}+G_{22}^{(00)}+G_{12}^{(00)}+G_{21}^{(00)}  } \, .
 \end{eqnarray}
\end{subequations}
in terms of the coefficients $X_{ij}^{(00)}$ ($X=G, B, \Gamma, \Xi$)
defined by Eqs.~(\ref{eq:slin}) and (\ref{scatt0sc}). 

It is then straightforward to see that the reciprocity relations read
specifically
\begin{subequations}
\begin{eqnarray}
G(\mathcal{H})& =& G(-\mathcal{H}) \, , \\
\Xi(\mathcal{H}) &=& \Xi(-\mathcal{H}) \, , \\ 
\label{eq:seebeck_peltier_sc_2}
B(\mathcal{H})\; T_0 &=& \Gamma(-\mathcal{H}) \, .
\end{eqnarray}
\end{subequations}
In particular the presence of superconductivity forces one to invert the sign of the TRS
breaking field in the relation of Eq.~(\ref{eq:seebeck_peltier_sc_2}) between Seebeck and
Peltier coefficients.

\subsection{Symmetry of the thermopower}
\label{Sect:symm-of-thermopower}

The symmetry of the two-terminal 
thermopower, $S=-B^{(00)} /G^{(00)} $ is not specified 
in the presence of superconductivity~\cite{Jac10}. 
The Seebeck coefficients read
\begin{eqnarray}
B^{(00)}_{ij} ({\cal H})&=& \frac{2e}{hT_0} \int_0^\infty {\rm d} \veps \,(- \partial_\veps f ) \, \veps \, 
\nonumber \\
& & \qquad  \times
\big[{\mathcal T}_{ij}^{(ee;00)}(\veps,{\cal H})+{\mathcal T}_{ij}^{(eh;00)}(\veps,{\cal H})
\nonumber \\
& & \qquad  \quad 
-{\mathcal T}_{ij}^{(he;00)}(\veps,{\cal H})-{\mathcal T}_{ij}^{(hh;00)}(\veps,{\cal H}) \big] \, .
\qquad
\end{eqnarray}
From this expression we see that  thermoelectric effects 
vanish, $B^{(00)}_{ij} = 0$, if PHS is enforced;
we thus consider this equation in the absence of PHS.
From Eq.~(\ref{eq:smatr_lms}) we know that 
${\mathcal T}_{ii}^{(\mu\mu;00)}(\veps,{\cal H}) = {\mathcal T}_{ii}^{(\mu\mu;00)}(\veps,-{\cal H})$, while
${\mathcal T}_{ii}^{(eh;00)}(\veps,{\cal H}) = {\mathcal T}_{ii}^{(he;00)}(\veps,-{\cal H})$. Together with
unitarity, $\sum_{j,\nu} {\mathcal T}_{ij}^{\mu \nu}(\veps,{\cal H}) = N_i^\mu$ and assuming that
the number $N_j^\mu$ of transport channels depends neither on the quasiparticle type nor on the
magnetic field, we readily obtain that $B^{(00)}_{ij}({\cal H}) = B^{(00)}_{\rm even}({\cal H}) + 
B^{(00)}_{\rm odd}({\cal H})$ is the sum of an even and an odd component,
\begin{subequations}\label{eq:evenodd}
\begin{eqnarray}
B^{(00)}_{\rm even}({\cal H}) &=&  \frac{2e}{hT_0} \int_0^\infty {\rm d} \veps \, (- \partial_\veps f ) \, \veps \, 
\nonumber \\
& & \times
\big[{\mathcal T}_{ij}^{(ee;00)}(\veps,{\cal H})-{\mathcal T}_{ij}^{(hh;00)}(\veps,{\cal H}) \big]\, , \qquad \quad
\\
B^{(00)}_{\rm odd}({\cal H}) &=&\frac{2e}{hT_0} \int_0^\infty {\rm d} \veps \, (- \partial_\veps f ) \, \veps \, 
\nonumber \\
& & \times
\big[ {\mathcal T}_{ij}^{(eh;00)}(\veps,{\cal H})-{\mathcal T}_{ij}^{(he;00)}(\veps,{\cal H}) \big] \, .
\end{eqnarray}
\end{subequations}
where $B^{(00)}_{\rm even}(-{\cal H}) =B^{(00)}_{\rm even}({\cal H})$
and $B^{(00)}_{\rm odd}(-{\cal H}) =-B^{(00)}_{\rm odd}({\cal H})$.
In the absence of Andreev scattering, $B^{(00)}({\cal H}) = B^{(00)}_{\rm even}({\cal H})$ is strictly
even in two-terminal geometries, however Andreev scattering gives rise to an odd component.
The asymmetric Andreev interferometers considered in Ref.~[\onlinecite{Jac10}] were devised to 
render $B^{(00)}_{\rm odd}({\cal H})$ finite on mesoscopic average, which led to 
an antisymmetric thermopower in such systems. There are currently no known hybrid systems
which have a finite-average $B^{(00)}_{\rm even}({\cal H})$. 
Recent theoretical works pointed out asymmetries in the thermopower of metallic
systems in the presence of inelastic scattering, which is of interest because asymmetric
thermopower may lead to more efficient thermal engines~\cite{Sai11,San11}. Hybrid
systems are examples of systems with purely elastic scattering and antisymmetric thermopower.

\begin{figure*}
\includegraphics[width=14.5cm]{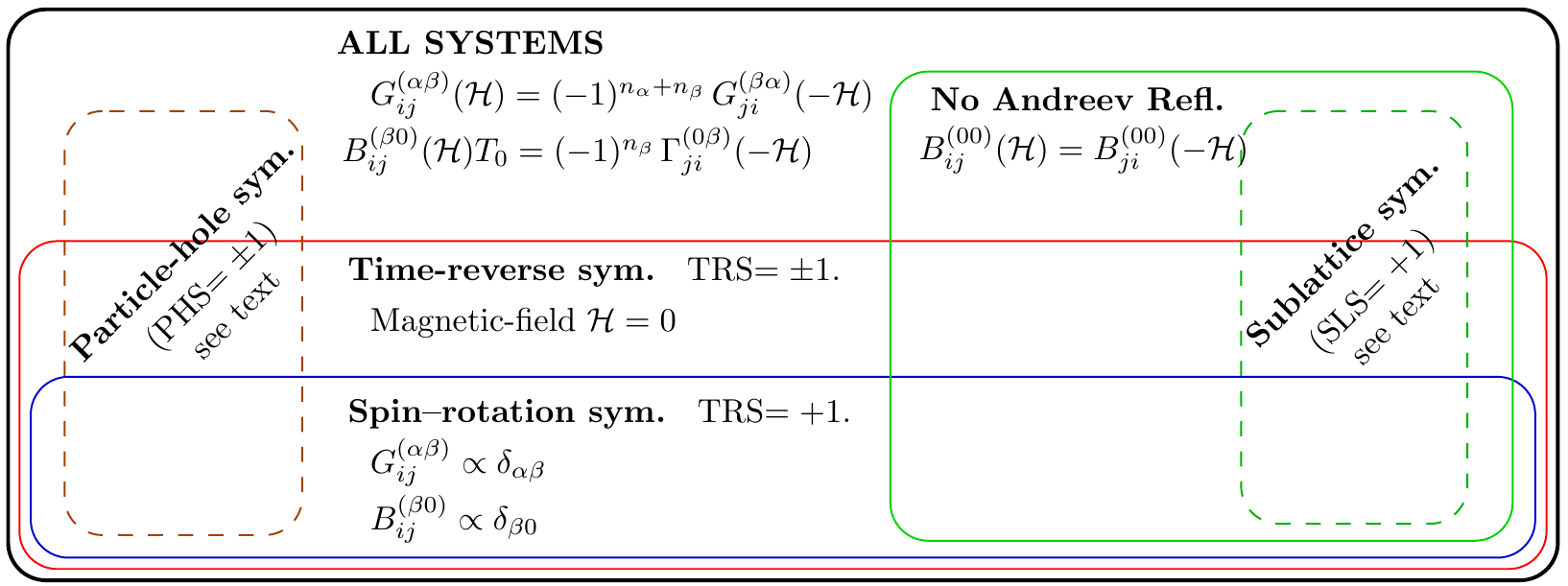}
\caption{\label{fig:general-results} 
Venn diagram summarizing the Onsager relations we derived. Here $\al,\be \in \{0,x,y,z\}$, 
$n_0=0$ and $n_x=n_y=n_z=1$.
Any system obeys the relations given in all boxes it is within. For instance
the Onsager relations for a generic system with time-reversal symmetry (TRS)
read $G_{ij}^{(\al\be)}= (-1)^{n_\al+n_\beta}G_{ji}^{(\be\al)}$ and 
$B_{ij}^{(\be 0)}\,T_0= (-1)^{n_\beta}\Gamma_{ji}^{(0\be)}$,
since TRS requires that any external magnetic field ${\cal H}=0$.
If that system also had SRS, then the only nonzero coefficients are those with repeated upper-indices ($\alpha=\beta$ for $G$
and $\beta=0$ for $B$) for which
$G_{ij}^{(\beta\beta)}=G_{ji}^{(\beta\beta)}$ and $B_{ij}^{(00)}\,T_0= \Gamma_{ji}^{(00)}$.
If the system contains no Andreev reflection (i.e.\ no superconductors), 
one additionally has~\cite{But90} $B_{ij}^{(00)} ({\cal H}) =B_{ji}^{(0 0)} (-{\cal H})$. 
In the case of PHS and SLS, the extra symmetry relations derived in the text are 
summarized in Table~\ref{table5}.}
\end{figure*}

\subsection{Onset of thermoelectric effects upon breaking of PHS}  
\label{sec:breakingphs}

Thermoelectric effects vanish identically in all Altland-Zirnbauer symmetry classes because of 
PHS. However in physical systems PHS is often at least partially broken, 
leading to finite thermoelectric effects. Here we show that the symmetry of such thermoelectric 
effects is subtly dependent on {\it how} PHS symmetry is broken.

To that end we consider the Andreev interferometer shown in Fig.~\ref{Fig:SC}. 
A two-terminal chaotic ballistic or disordered diffusive quantum dot is connected to a 
superconducting loop via
two contacts. The superconducting phase difference at the two contacts can be tuned by a magnetic
flux piercing the loop. There are two important time scales in the system, (i) the typical time 
$\tau_{\rm Andr}$ between two consecutive Andreev reflections at the superconducting contact, and 
(ii) the escape time $\tau_{\rm esc}$ to one of the normal leads. We additionally choose a special
geometry where the average time to reach one of the two superconducting 
contacts from one of the normal leads is longer -- this is achieved by an extra ballistic "neck" of length
$\ell$ between the cavity and the superconducting contact (see Fig.~\ref{Fig:SC}). Because of the neck,
quasiparticles need
an additional time delay $\delta \tau = \ell/v_{\rm F}$ to reach the left superconducting contact
from a normal lead. Together with this time delay,
a magnetic flux piercing the superconducting loop and making the superconducting
phase difference $\phi_{\rm sc}$ finite also breaks PHS, thereby turning thermoelectric effects
on~\cite{Jac10,Eng11}.

Formally, PHS requires that
$\tau_{\rm Andr} \to 0$,
which practically means that $\tau_{\rm Andr}$ has to be smaller than any
other time scale and any other inverse energy scale. 
When this is not the case,  
transport processes without any Andreev reflection exist, giving contributions to the conductance
that fluctuate randomly in energy around the Fermi energy. This 
breaks PHS
and leads for instance to finite, albeit relatively weak, thermopower~\cite{Lan98,God99}.
More generally, breaking PHS can be achieved in three different ways, 
\begin{itemize}
\item[(i)] rendering escape into the normal leads faster (for instance by widening the normal leads), 
until $\tau_{\rm esc} \sim \tau_{\rm Andr}$, \\[-6mm]
\item[(ii)] raising the temperature until $(k_{`\rm B}T)^{-1} \sim \tau_{\rm Andr}$, or \\[-6mm]
\item[(iii)] changing the flux through the superconducting
loop so that $\phi_{\rm SC} \neq 0,\pi$, when the neck length $\ell$ is finite. 
\end{itemize} 


In case (i), a significant proportion of quasiparticles go from one normal lead to another without Andreev reflection. Then contributions to $\mathcal{T}_{ij}^{(\mu\mu;\alpha\beta)}({\cal H})$ which arise from
processes without Andreev reflections will start to dominate thermoelectric transport, meaning
$B^{(00)}_{\rm even} \gg B^{(00)}_{\rm odd}$ [as defined in Eq.~(\ref{eq:evenodd})]. Thus, 
thermoelectric effects acquire the same symmetry as systems without SC contacts, i.e. 
they become predominantly even.
  
The situation is more complicated in case (ii), where
both $\mathcal{T}_{ij}^{(\mu\mu;\alpha\beta)}({\cal H})$ and 
$\mathcal{T}_{ij}^{(\mu\overline{\mu};\alpha\beta)}({\cal H})$ have similar magnitude.
In the absence of a neck, $\ell=0$, thermoelectric effects vanish on average and are
dominated by mesoscopic fluctuations~\cite{Jac10,Eng11}.
An analysis of these mesoscopic fluctuations analogous to that in 
Ref.~[\onlinecite{Jac10}] shows that
there is no correlation between $B_{ij}^{(00)}({\cal H})$ and $B_{ji}^{(00)}(-{\cal H})$, 
so that the thermoelectric effects have no particular symmetry
beyond the generic Onsager reciprocities given in Table~\ref{table3}.
In particular, for a two terminal device 
$B^{(00)}_{\rm even}$ and $B^{(00)}_{\rm odd}$ are independent random variables
with the same variance. Thus
for a given Andreev interferometer (given disorder or cavity shape) 
either quantity could be positive or negative,
and either could have a larger magnitude than the other.  

Finally in case (iii), the physics changes completely. Due to the presence of a finite-sized neck,
$\ell \ne 0$, and superconducting phase difference $\phi_{\rm sc} \neq 0,\pi$,
the system develops a large average thermopower which is
an odd function of the flux $\phi_{\rm sc}$~\cite{Vir04,Tit08,Jac10,Eng11}, 
with a much smaller even component coming from mesoscopic fluctuations \cite{Jac10}.  

In summary depending on how particle-hole symmetry is broken, one gets a thermopower
which is predominantly even in ${\cal H}$ [case (i)], predominantly odd in ${\cal H}$ [case (iii)],
or which has no particular symmetry [case (ii)].

\section{Conclusions}\label{sec:conclusion}

We have derived a complete list of reciprocity relations for coupled electric, spin, thermoelectric and
spin caloritronic transport effects  in all ten symmetry classes for single-particle Hamiltonian systems. 
Several of these 
relations appeared in one way or another in earlier works, and the main novelties we found are
(i) reciprocities in spintronics and spin caloritronics pick a number of additional minus signs reflecting
spin current injection and measurement, (ii) a number of special relations have been listed in 
Table~\ref{table5}, which exist only in specific symmetry classes, (iii) 
we clarified the exact form of Onsager relations in the presence of superconductivity, and
(iv) we derived all Onsager relations for transport in spintronics and spin caloritronics in the
presence of superconductivity.
We present a pictorial summary of the 
Onsager reciprocity relations we derived in Fig.~\ref{fig:general-results}.

Generally speaking, our investigations of the specific reciprocities shown in Table~\ref{table5}
allowed us to clarify the form that the Seebeck-Peltier relations take in the presence of superconductivity. 
While the two relations,
$B_{ij}^{(00)}({\cal H}) T_0=\Gamma_{ij}^{(00)}({\cal H})$ and 
$B_{ij}^{(00)}({\cal H}) T_0=\Gamma_{ji}^{(00)}(-{\cal H})$ exist  in 
purely metallic systems, only one of these two Onsager relations 
survives in the presence of superconductivity,
that being $B_{ij}^{(00)}({\cal H}) T_0=\Gamma_{ji}^{(00)}(-{\cal H})$.

\section*{Acknowledgments}

We thank J. Li for discussions and useful comments on the manuscript.
This work was supported by the Swiss Center of Excellence MANEP, the European
STREP Network {\it Nanopower} and the NSF under grant PHY-1001017.


\begin{thebibliography}{}
\bibitem{caveat1} Care should be taken when considering effective Hamiltonian
formulations for systems with Majorana bound states, whose presence usually
breaks TRS and should thus be incorporated one way or another into the field
${\cal H}$. This ambiguity does not exist when considering microscopic Hamiltonians.
\bibitem{footnote:TRS-plus-minus-one}
For TRS, whether the antiunitary operator squares to $+1$ or $-1$ has physical meaning.
It comes from the sign of the leading order quantum correction
to conductance, with 
$+1$ leading to a weak-localization term and $-1$ leading to a 
weak-antilocalization term. See sect.~VII of  Ref.~[\onlinecite{alt97}].

\bi{caveat2} 
For an $s$-wave superconducting order parameter, PHS reduces to $P=-i \tau^{(y)} K$ 
when there is SRS in the normal-metallic part of the system.
\end{thebibliography}
\end{document}